\documentclass{iaus}
\usepackage{graphicx}

\title[Dark stars in the dark ages] %% give here short title %%
{Effects of dark matter annihilation on the first stars}

\author[F.~Iocco et al.]   %% give here short author list %%
{F.~Iocco$^1$, A.~Bressan$^2$$^,$$^3$, E.~Ripamonti$^4$, R.~Schneider$^1$, A.~Ferrara$^3$, P.~Marigo$^5$}

\affiliation{$^1$INAF/Oss.Astr.di Arcetri; Largo E.~Fermi 5, 50125 Firenze, Italy\\[\affilskip]
$^2$INAF/Osservatorio Astronomico di Padova; Vicolo dell'Osservatorio 5, Padova, Italy\\[\affilskip]
$^{3}$SISSA; Via Beirut 4, Trieste, Italy\\[\affilskip]
$^{4}$Universit\`a degli Studi dell'Insubria, Dip. di Scienze Chimiche, Fisiche e Naturali; Via Valleggio 12, Como, Italy\\[\affilskip]
$^{5}$Universit\`a degli Studi di Padova, Dip. di Astronomia; Vicolo dell'Osservatorio 3, Padova, Italy}
\pubyear{2008}
\volume{255}  %% insert here IAU Symposium No.
\pagerange{1--5}
\jname{Low-Metallicity Star Formation: From the First Stars to Dwarf Galaxies}
\editors{L.K. Hunt, S. Madden \& R. Schneider, eds.}
\begin{document}

\maketitle

\begin{abstract}
We study the evolution of the first stars in the universe (Population III) from the early pre--Main Sequence (MS) until the end of helium burning in the presence of WIMP dark matter annihilation inside the stellar structure. The two different mechanisms that can provide this energy source are the contemporary contraction of baryons and dark matter, and the capture of WIMPs by scattering off the gas with subsequent accumulation inside the star.
We find that the first mechanism can generate an equilibrium phase, previously known as a {\it dark star}, which is transient and present in the very early stages of pre--MS evolution. The mechanism of scattering and capture acts later, and can support the star virtually forever, depending on environmental characteristic of the dark matter halo and on the specific WIMP model. 
\end{abstract}

\firstsection % if your document starts with a section,
\section{Introduction}
%####
Within the scenario of a $\Lambda$CDM cosmology, it has been recently recognized that if the dark matter (DM) dominant component is a weakly interacting massive particle (WIMP), its annihilation could play a relevant role on the formation and evolution of the first baryonic objects in our universe. Spolyar, Freese, and Gondolo (2008) noticed that during the proto--stellar phase, the cooling of the baryonic gas could be overcome by the energy deposition following the annihilation of DM concentrated in the star formation site. This is a consequence of the peculiar formation characteristics of the first stars, at the center of a minihalo whose gas cooling is dominated by the little efficient primordial chemistry; the authors suggested that this phase (called a {\it dark star}) could prevent the formation of the first stars, and be a new phase of stellar evolution.
Iocco (2008) and Freese, Spolyar, and Aguirre (2008) noticed that, if a star does eventually form, the process of WIMP capture by scattering could be so efficient that their subsequent accumulation and annihilation inside the celestial object may provide an energy source comparable or even exceeding its nuclear luminosity.
Motivated by these works, Iocco et al. (2008) and Freese et al. (2008b,c) have studied the early pre--MS phase, in which the baryonic structure is sustained by the annihilation of the DM accreted inside it by gravitational contraction, with the help of numerical codes. Both the groups find this phase is transient, although the techniques adopted are different and the details of the treatment lead to different duration estimates; they both conclude, however, that the collapse must continue at the end of this process, which for the sake of simplicity we call the Adiabatic Contraction ({\it AC}) phase.
In Iocco et al. (2008) (hereafter, I08), we have also studied the pre--MS phase of these stars in presence of annihilating scattered and captured DM ({\it SC} phase) and followed the evolution of stellar models of different mass until the end of the helium burning, in different DM environments. We find the duration of the MS is dramatically prolonged, up to a potentially everlasting phase (depending  on the choice of parameters, see later): the energy released by the annihilating DM can support the core at temperatures low enough that nuclear reactions are never ignited. Yoon, Iocco, and Akiyama (2008) and Taoso et al. (2008) also studied the {\it SC} phase, confirming our results with different codes and carrying their analysis further.
These proceedings are based on the results obtained by I08, to which we address the reader for detailed referencing and more quantitative details: here we aim to a more qualitative description of the physical processes at the basis of this class of objects.
%####
\section{Adiabatic Contraction phase}
%####
Early stars are thought to form in halos of M$_h\sim$10$^6$ M$_\odot$ and virial temperature T$\lesssim$10$^4$ K at redshift z$\sim$20.
The primordial, metal free composition of the gas, the absence of strong magnetic fields and turbulence make early star formation very different from the one in the older universe. Simulations tell us that the result of such peculiar environment are massive (30-300 M$_\odot$) stars that form in the very center of the halo (accurate review and referencing in 
the First Stars III proceedings, 2008).
The collapse of the baryonic material ``pulls'' also the collisional DM towards the center, thus contributing to the build--up of a central ``spike'' of DM; 
%similarly to what is thought to have happened at our Galactic Center at the time of its formation (Ullio, Zhao, and Kamionkowski 2001).
if that is indeed made of self-annihilating particles (as WIMP neutralinos are), its higher concentration in the center of the halo causes a huge enhancement of the annihilation term:
\begin{equation}
\frac{dL_{\rm DM}}{dV}=fc^2\frac{\rho^2\langle\sigma v\rangle}{m_\chi};
\label{FSDMannih}
\end{equation}
\noindent
with $\rho$ the local DM density, $m_\chi$ the neutralino mass --which we take to be 100GeV--, $\langle\sigma v\rangle$ the thermally--averaged 
annihilation rate. 
It is worth noting that this energy term depends on the self--annihilation rate,
whose value can be quite safely established through cosmological arguments as $\langle \sigma v \rangle = 3 \times 10^{-26}$~cm$^3$~s$^{-1}$, see for instance a recent review on particle DM (e.g. Bertone et al. 2005).
%The energy released by this annihilation can be injected in the form of different primaries and subsequent showers, depending on the type of WIMP particle; 
A fraction ($1-f$) of the whole energy is emitted in the form of neutrinos, and therefore lost by the system, whereas all the other products of annihilation can quickly thermalize inside the gas already at densities $n\sim$10$^{13}$\#/cm$^3$, as shown by Spolyar et al (2008); we take $f\sim$2/3, a typical value for a neutralino annihilation.
In order to study this peculiar object we have modified the {\it Padova} Stellar Evolution code to take into account  the energy released by DM annihilation. The initial models have been prepared as follows: $(i)$ the ``stellar'' profile (namely the baryonic structure) has been obtained by ``pumping'' a Zero Age Main Sequence star with an artificial energy source (as much as permitted by the stability of our numerical code);  for a 100M$_\odot$ star,  the resulting structure is an object of radius R$_\ast$=1.2$\times$10$^{14}$cm and effective temperature $T\sim$5$\times$10$^3$~K; $(ii)$ the DM profile has been obtained in the approximation of an adiabatically contracted DM profile (from an original NFW) which has been matched to the baryonic structure in the center, which dictates the gravitational potential. For details on the characteristics of the halo and on the adiabatic contraction approximation used we address the reader to I08.
This adiabatically contracted DM profile represents the initial model, at the time we start our analysis:
we have implemented in our code a routine which allows to follow the adiabatic contraction of the DM inside our stellar object, account for the energy released by its annihilation, and so self--consistently describe the evolution of the DM coupled to the baryons.
In this way we can follow the contraction of the protostar toward the Main Sequence including all relevant energy sources (gravitational, DM annihilation, and possibly nuclear) and adopt a complete equation of state for the gas.

\begin{figure}[!t]
% \vspace*{-2.0 cm}
\begin{center}
\begin{tabular}{cc}
\includegraphics[angle=90,width=2.4in]{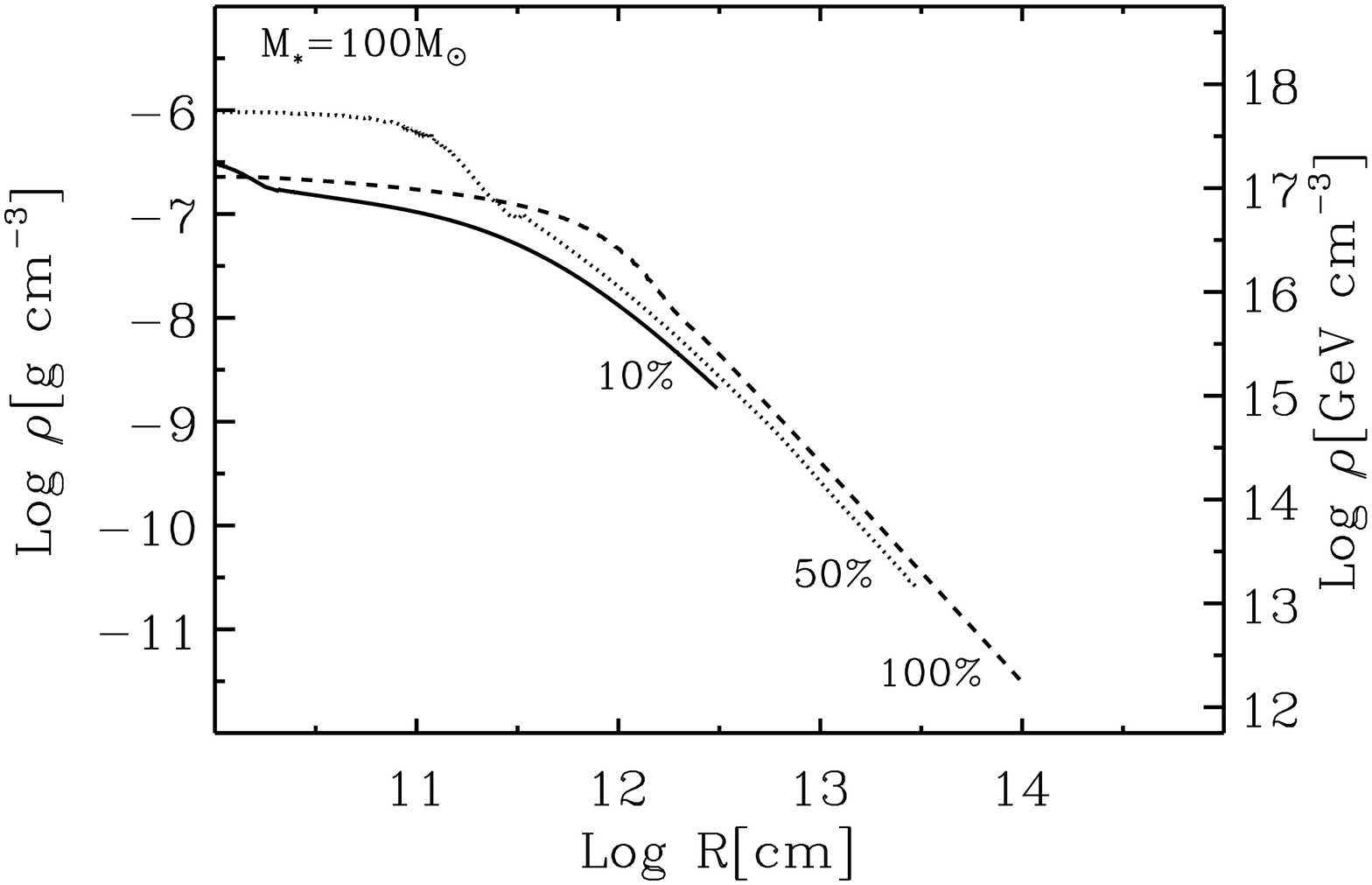} \ \ \ \ \ \ \
\includegraphics[angle=90,width=2.4in]{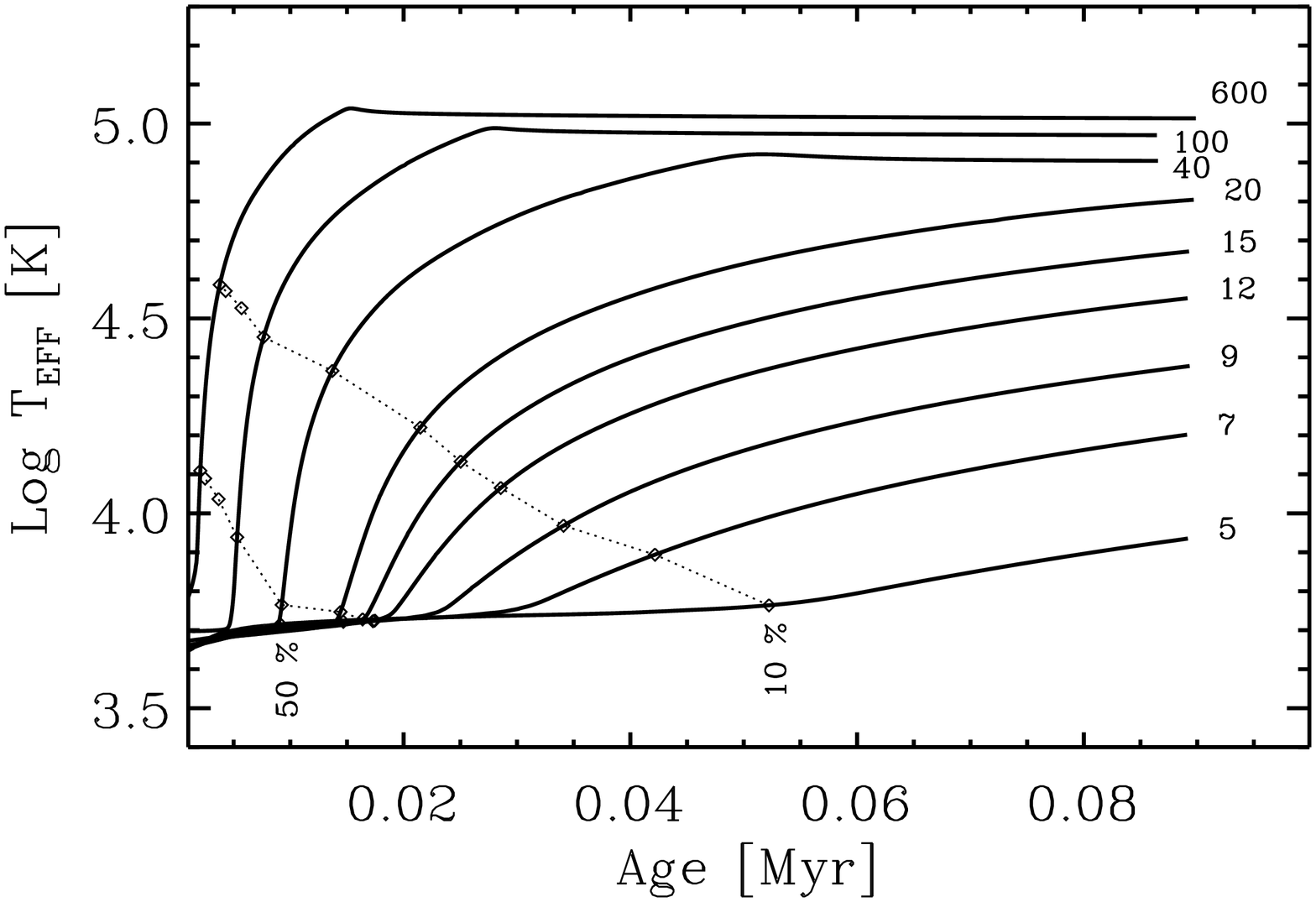}
\end{tabular}
\caption{In the left panel: DM density profiles, truncated at the stellar radius, for a 100M$_\odot$ at different times during the {\it AC phase}.  In the right panel: evolution of the effective temperature for all stellar models, benchmarked by the relative contribution of DM annihilation to the total luminosity of the star. See text for details}
\label{fig:ACphase} 
\end{center}
\end{figure}

For the adopted AC parameters, DM annihilation is able to support the baryonic structure in the very early phases of  gravitational contraction, in all our selected stellar models (5--600 M$_\odot$)
The protostars contract "pulling in" more and more DM until an equilibrium is reached in very short times, ${\cal O}$(10$^2$yr).

%At this stage, all the stellar masses we have followed (5, 7, 9, 12, 15, 20, 40, 100, 200, 400, 600 M$_\odot$) are unstable (nuclear reactions are inefficient for the low temperature of the core T$_c\sim$10$^5$~K) and contract, ``pulling in'' more DM and finding an equilibrium in very short times ${\cal O}$(10$^2$yr): DM annihilation supports the baryonic structure against gravitational collapse.
%However, this phase is doomed to be transient: the ``erosion'' of the DM profile as a consequence of the DM annihilation implies an out--of--equilibrium phase, the contraction of the baryonic structure, and of the DM pulled in together with it.
%If the new DM configuration is able to sustain the baryonic structure, there will be another equilibrium transient, followed by another contraction and so on, till the contraction of DM will not be fast enough to build an ``efficient'' cusp. At this point the object becomes unstable and keeps on contracting, following the evolution of a pre--MS star.
%and eventually igniting the nuclear reactions at the Zero Age Main Sequence point on the HR diagram.
However, this phase is doomed to be transient.
As the annihilation proceeds, the star contracts because of the shrink  of DM luminosity caused by its consumption.
But this further contraction is not able to restore the previous DM density profile, and the star finds its equilibrium at a lower luminosity.
In this phase the star descends slowly along the Hayashi line.
Besides consumption due to annihilation, the star continuosly {\sl looses} the external shells of DM, 
because of the different response af baryons and DM to the growing gravitational potential.
Eventually  the contraction of DM becomes not fast enough to restore an ``efficient'' cusp and the DM luminosity is no more able to balance the stellar energy losses.
At this point the star  keeps on contracting on a typical the Kelvin-Helmholtz time-scale,
terminating the AC phase and moving toward the main sequence.
%The instability of this phase is a consequence of different timescales: the Kelvin-Helmholtz time, typical of the contraction process for the star, and the time the DM cusp takes to re--build a ``energetically efficient'' profile. 
%minicusp peaked enough to balance
By defining the duration of this phase $\tau_{AC}$ as the the time needed from the DM annihilation to scale from 100\% to 50\% of the total luminosity  of the object,
we find typical values of order $\tau_{\rm AC}\sim$10$^3$yr, ranging from $\tau_{\rm AC}$=2.1$\times$10$^3$yr for a 600$M_\odot$ star up to $\tau_{\rm AC}$=1.8$\times$10$^4$yr for a 9$M_\odot$ star.
%The rate of gravitational energy release is larger in more massive stars, so it comes as no surprise that the stalling phase is shorter for higher mass stars.
In Figure \ref{fig:ACphase} we show the effective temperature evolution for different stellar models and the DM profile {\it inside} the baryonic structure for a 100M$_\odot$ star at different times during the stellar collapse.
The benchmark values correspond to the fractional contribution of DM annihilation to the total luminosity.
%#####
\section{Scattering and Capture phase}
%#####
The other physical mechanism able to concentrate DM inside a star is 
the capture of WIMPs by means of elastic scattering with the gas
particles that constitute the object. The captured WIMPs thermalize with the gas
and eventually reach a thermally relaxed state inside the star: the density profile $n_\chi$ is
\begin{equation}
n_{\chi}(R) =n^c_\chi\exp(-R^2/r_\chi^2), \ \  n_\chi^c=\frac{C_\ast \tau_\chi}{\pi^{3/2}r^3_\chi}, \ \ 
r_{\chi}=c\left(\frac{3kT_c}{2\pi G\rho_c m_\chi}\right)^{1/2};
\label{DMprofile} 
\end{equation}
\noindent
where $T_c$ and $\rho_c$ are the stellar core temperature and density, respectively, 
and $C_\ast$ the WIMP capture rate, discussed in the following. 
It can be seen that $r_\chi$ is in general much smaller than the star core,
$r_\chi\sim$10$^9$~cm for a 100M$_\odot$ star.
For each volume element, the DM annihilation energy released is given by
Equation \ref{FSDMannih}; however, capture and annihilation reach the equilibrium on
timescales much shorter than the stellar lifetime (see Iocco (2008) and I08 for a more detailed discussion
of this issue), and the amount of energy released inside the star, the ``dark luminosity''
$L_{DM}$ reads $L_{DM}$=$fm_\chi C_\ast$, being dictated by the capture rate, $C_\ast$. 
The latter has the characteristic of a scattering term, and 
can therefore be cast as:
\begin{equation}
C_\ast \propto   M_\ast v_{\rm esc}^2 \frac{\sigma_0\rho}{m_\chi}.
\label{propcap}
\end{equation}
\noindent
with $\sigma_0$ the elastic scattering cross section between WIMP and baryons,
$\rho$ the DM density {\it outside} the star, $v_{esc}$ the escape velocity at the 
surface of the star, and the subscript $\ast$ referring to stellar quantities throughout this
paper.
We refer to I08 and references therein for a more quantitative discussion of the 
capture rate, but we wish here to stress on few peculiarities of the {\it SC} process.
The energy is indeed provided by means of DM {\it annihilation} in the bosom of the star;
however, the bottleneck of the capture/scattering mechanism is given by capture: the star
cannot burn more WIMPs than it accretes. Therefore, at the equilibrium, the annihilation
is dictated by the capture and shows the parameter dependence of a scattering process.
Also, this process is sensitive to the WIMPs that stream through the star, and thus the halo plays the role
of a ``reservoir'', virtually unexhaustable\footnote{The DM mass-energy content of the halo is much bigger 
than the energy used by the star: only approximately 10M$_\odot$=10$^{-5}$M$_h$ are entirely converted 
into energy in 10Gyr, at a rate 10$^{39}$erg/s. Only 0.1M$_\odot$ of DM can support a 100M$_\odot$ star 
for ten times longer than its natural lifetime, 10$^6$yr.}.
It depends on $\sigma_0$ rather than $\langle\sigma v\rangle$, and {\it linearly} on the {\it environmental} 
DM density $\rho$. 
%Typical timescales for the equilibrium to be reached are reported in Iocco (2008), and are of order 1yr at the most, for an object close to ZAMS, so it can be considered as istantaneous during the MS.
However, when the object is in the {\it AC} phase, the {\it SC} mechanism is playing little role because of the small capture rate (the star is an inefficient ``net'' for capturing DM) and of the very long
equilibrium timescale at these stages, much longer than the Kelvin-Helmholtz time. By contracting, the 
star enhances the efficiency of the {\it SC} process and DM annihilation contributes to the stellar luminosity.

\begin{figure}[!t]
% \vspace*{-2.0 cm}
\begin{center}
\begin{tabular}{cc}
\includegraphics[angle=90,width=2.4in]{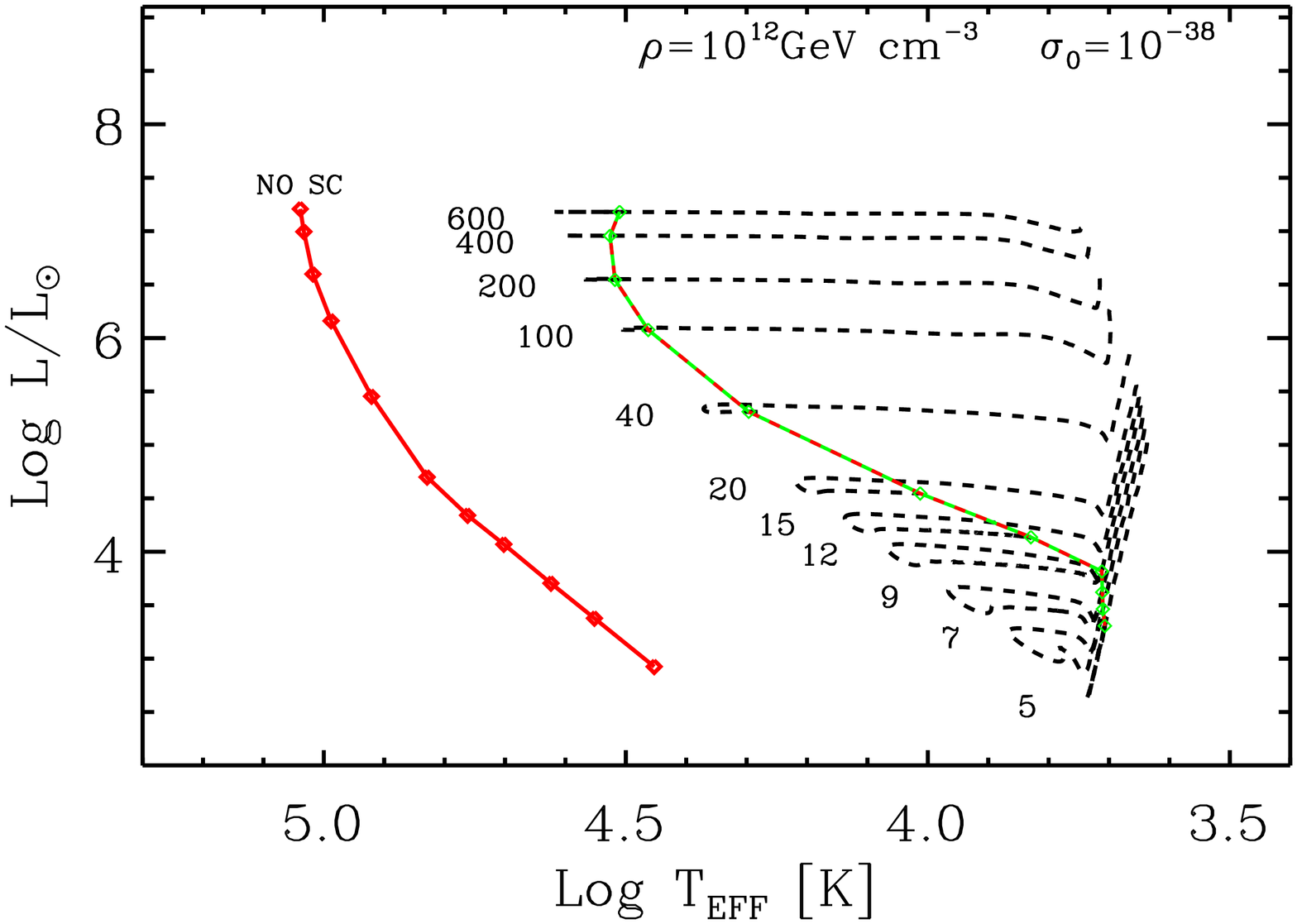} \ \ \ \ \ 
\includegraphics[angle=90,width=2.4in]{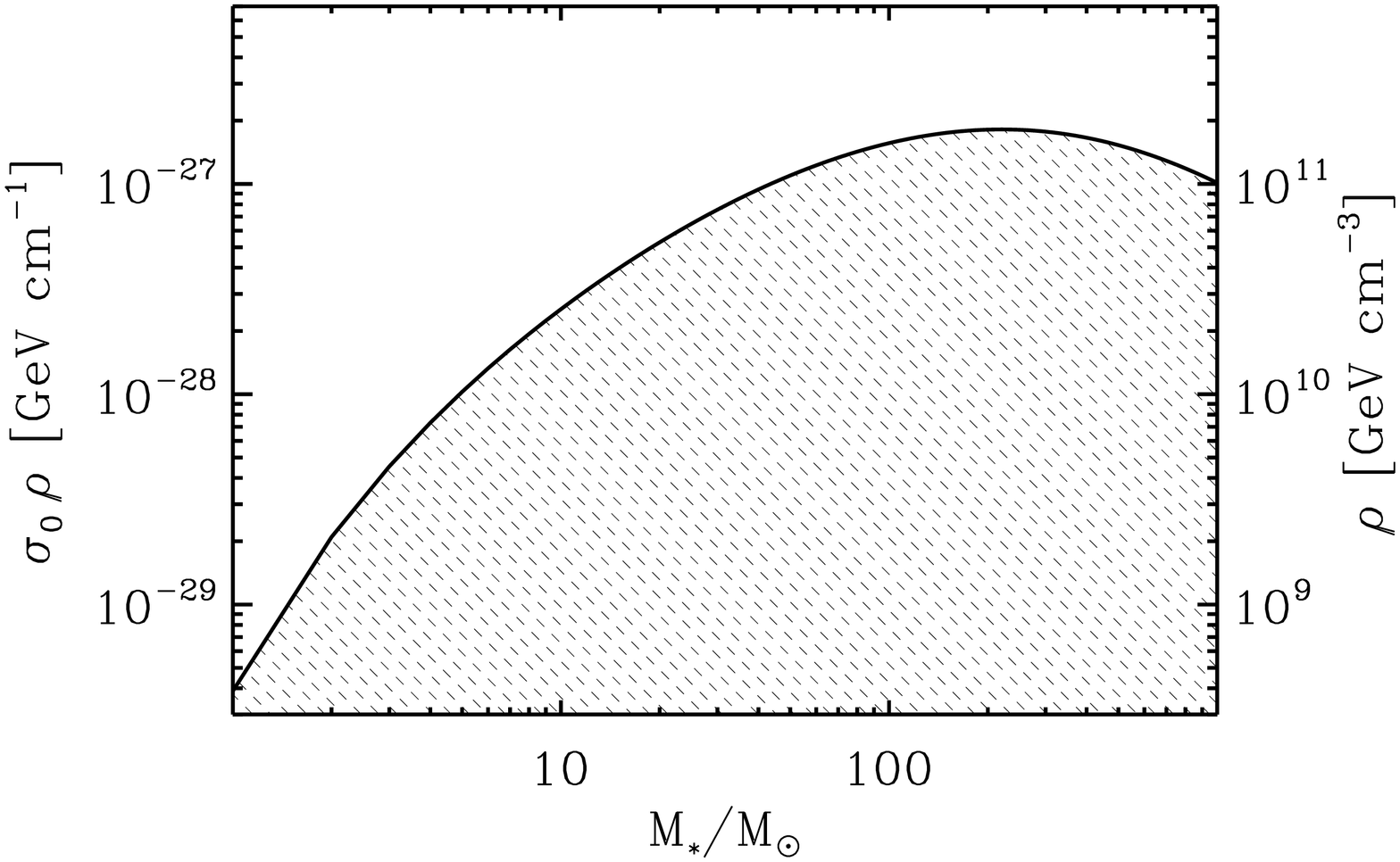}
\end{tabular}
\caption{In the left panel: HR diagram for metal--free stars in presence of a DM environmental density $\rho$=10$^{12}$GeV, compared to their normal position in the HR; sold curve labeled {\it NO SC}. For all masses, the ZAMS is never reached, as stars are entirely supported by DM annihilation. In the left panel: stars in the white region, above the solid line never ignite nuclear reactions, being ``frozen'' in their evolution. In the shaded area they do evolve at different rates, being only partially supported by DM burning.
See text for details.}
\label{fig:SCphase} 
\end{center}
\end{figure}

For $\sigma_0$=10$^{-38}$ (at the level of the current upper limit for the
spin--dependent elastic scattering cross section (Desai et al., 2004) ), and an environment 
density $\rho$=10$^{12}$GeV/cm$^3$ (a likely value achieved around 
the star, see Figure \ref{fig:ACphase}), the energy provided by the DM annihilation inside the star is able to entirely support all our models before they get to the ZAMS, therefore not igniting nuclear reactions.
Their locus on the HR diagram can be observed in Figure \ref{fig:SCphase}, compared 
with their ZAMS one. 
Lower values of $\rho$ or $\sigma_0$ make the DM contribution smaller,
and the star can contract increasingly, thus progressively enabling the ignition
of nuclear reactions. Stars in the white region of the right panel of Figure \ref{fig:ACphase}
do never ignite nuclear reactions; stars below the solid curve, in the shaded area do.
The closer to the curve, the longer their lifetimes will be, as energy contribution required by
nuclear burning is lower, and their chemical evolution is slower, see I08.

\section{Conclusions}
We have studied the effects of WIMP dark matter annihilation from the late stages of stellar formation through the end of helium burning.
Addressing the reader to Iocco et al. (2008) for quantitative details, here
we wish to stress the fundamental physical difference between the {\it SC} mechanism,
which needs a weak process to mediate the capture and thermalization of
DM particles (and therefore the DM density profile inside the star), and the
{\it AC} one, where the DM density profile is dictated by gravitational contraction
only.
The difference in the two physical processes dictates diverse
characteristics for the two phases: the {\it AC} is transient, and takes
place early in the pre--Main Sequence evolution of the star, when it is almost
at the proto--stellar stage. 
The {\it SC} phase becomes active when the star is
at the bottom of the Hayashi track or later in the pre--MS evolution, and it 
can dramatically extend the stellar life, up to orders of magnitude more than 
its standard lifespan, depending on parameters.

As a note added in proof, we address the reader to a paper very similar to this one, Freese et al. (2008d) in these proceedings; and also to the analysis of Natarajan, Tan, and O' Shea (2008) who find results in good agreement with Spolyar et al. 's, studying gas and dark matter profiles from cosmological simulations of first star formation.
%The {\it SC} phase is ``repeatable'', meaning that a normal star which is put in high enough DM concentration can become a DM burner again. This has been found by Fairbairn, Scott, and Edsjo (2008), who ``feed'' ZAMS stars with increasing DM densities and observe them expand and cool down towards the Hayashi track.
%With some {\it caveats}, this is not the case for the {\it AC} phase, which relies on the contraction of the baryonic cloud for the build up of the cusp.


\begin{thebibliography}{}

%%\cite{Bertone:2004pz}
\bibitem[Bertone et al. 2005]{Bertone:2004pz}
Bertone G., Hooper D., Silk J., 2005, Phys. Rep. 405, 279

\bibitem[SuperKamiokande 2004]{Desai:2004pq}
 Desai S. {\it et al.}  [Super-Kamiokande Collaboration], 2004, Phys. Rev. D, 70, 109901

%\bibitem[Fairbairn et al. 2008]{Fairbairn:2007bn}
%Fairbairn M., Scott P., Edsjo J., 2008, Phys. Rev. D, 77, 047301

\bibitem[First Stars III proceedings (2008)]{FS3proc} 
"First Stars III", 2008, AIP Conf.~Proc., 990, T.~Abel, A.~Heger, and B.~W.~O'.~Shea eds 

%\cite{Freese:2008ur}
\bibitem[Freese et al. (2008a)]{Freese:2008ur}
 Freese K., Spolyar D., Aguirre A.,
  arXiv:0802.1724 [astro-ph].

%\cite{Freese:2008hb}
\bibitem[Freese et al. (2008b)]{Freese:2008hb}
Freese K., Gondolo P., Sellwood J. A., Spolyar D., [Freese et al. (2008b)]
  arXiv:0805.3540 [astro-ph].

%\cite{Freese:2008wh}
\bibitem[Freese et al. (2008c)]{Freese:2008wh}
  Freese K., Bodenheimer P., Spolyar D., Gondolo P.,
  %``Stellar Structure of Dark Stars: a first phase of Stellar Evolution due to
  %Dark Matter Annihilation,''
  arXiv:0806.0617 [astro-ph].
  %%CITATION = ARXIV:0806.0617;%%

%\cite{Freese:2008be}
\bibitem[Freese et al. (2008d)]{Freese:2008be}
  Freese K., Spolyar D., Aguirre A., Bodenheimer P., Gondolo P., Sellwood J.~A., Yoshida N.,
  %``Dark Stars: Dark Matter in the First Stars leads to a New Phase of Stellar
  %Evolution,''
  arXiv:0808.0472 [astro-ph].
  %%CITATION = ARXIV:0808.0472;%%

\bibitem[Iocco 2008]{Iocco:2008xb}
%  [arXiv:0802.0941 [astro-ph]].
  %%CITATION = ASJOA,677,L1;%%
Iocco F., 2008, ApJ, 677, L1 
  
\bibitem[Iocco et al.~(2008)]{Iocco:2008rb} 
 Iocco F., Bressan A., Ripamonti E., Schneider R., Ferrara A., Marigo P.,  
  MNRAS in press, arXiv:0805.4016 [astro-ph].

%\cite{Natarajan:2008db}
\bibitem[Natarajan et al. (2008)]{Natarajan:2008db}
  Natarajan A., Tan J.~C., O'Shea B.~W.,
  %``Dark Matter Annihilation and Primordial Star Formation,''
  arXiv:0807.3769 [astro-ph].
  %%CITATION = ARXIV:0807.3769;%%

\bibitem[Yoon et al. `08]{Yoon:2008km}
  Yoon S.~C., Iocco F., Akiyama S.,
  arXiv:0806.2662 [astro-ph].

\bibitem[Spolyar et al `08]{Spolyar:2007qv}
Spolyar D., Freese K., Gondolo P., 2008, Phys. Rev. Lett., 100, 051101

\bibitem[Taoso et al. `08]{Taoso:2008kw}
 Taoso M., Bertone G., Meynet G., Ekstrom S.,
  arXiv:0806.2681 [astro-ph].

% \bibitem[Ullio et al. `01]{Ullio:2001fb}  
 %Ullio P., Zhao H. and Kamionkowski M., 2001  Phys. Rev. D  64, 043504 (2001)

\end{thebibliography}
\end{document}